\documentclass[twocolumn,aps,preprintnumbers,prl,floatfix,showpacs,amsmath,amsfonts,amssymb,floatfix,superscriptaddress,
reprint
]{revtex4-1}

\usepackage{graphics}
\usepackage{color}
\usepackage{float}

\usepackage{amssymb}
\usepackage{amsmath}
\usepackage{overpic}
\usepackage{epstopdf}

\usepackage{bm}
\usepackage{graphicx}

\newcommand{\lnsco}{La$_{1.48}$Nd$_{0.4}$Sr$_{0.12}$CuO$_{4}$}

\begin{document}

\title{Collective Dynamics and Strong Pinning Near the Onset of Charge Order in \lnsco}

\author{P.~G.~Baity}
\affiliation{National High Magnetic Field Laboratory and Department of Physics, Florida State University, Tallahassee, Florida 32306, USA}
\author{T. Sasagawa}
\affiliation{Materials and Structures Laboratory, Tokyo Institute of Technology, Kanagawa 226-8503, Japan}
\author{Dragana Popovi\'c}
\email{dragana@magnet.fsu.edu}
\affiliation{National High Magnetic Field Laboratory and Department of Physics, Florida State University, Tallahassee, Florida 32306, USA}

\date{\today}

\begin{abstract}
The dynamics of charge-ordered states is one of the key issues in underdoped cuprate high-temperature superconductors, but static short-range charge-order (CO) domains have been detected in almost all cuprates.  We probe the dynamics across the CO (and structural) transition in \lnsco\, by measuring nonequilibrium charge transport, or resistance $R$ as the system responds to a change in temperature and to an applied magnetic field.  We find evidence for metastable states, collective behavior, and criticality.  The collective dynamics in the critical regime indicates strong pinning by disorder.  Surprisingly, nonequilibrium effects, such as avalanches in $R$, are revealed only when the critical region is approached from the charge-ordered phase.  Our results on \lnsco\, provide the long-sought evidence for the fluctuating order across the CO transition, and also set important constraints on theories of dynamic stripes.
\end{abstract}

\maketitle

The role of various forms of charge and spin orders (``stripes'') observed in underdoped cuprate high-$T_c$ superconductors is one of the main open issues in the field \cite{Fradkin2015}.  In particular, the dynamics of charge-ordered states and the search for fluctuations of the incipient charge order (``fluctuating order'', or ``dynamic stripes'') have been the subject of intensive research with the goal to clarify their relationship to high-$T_c$ superconductivity \cite{Kivelson2003,MVojta2009}.  However, except in La$_{1.9}$Sr$_{0.1}$CuO$_4$ \cite{Gedik2013}, static COs have been found in all cuprates  \cite{Wise2008,Chang2012,LeB2012,Ghirin2012,Rosen2013,Wu2013,Blanco2013,Blackburn2013,Tacon2014,Comin2014,Wu2015,Chen2016}.  The remarkable stability of the CO and its short-range nature are usually believed to be due to the pinning by disorder \cite{Robertson2006,Maestro2006,Nie2014}, but it has also been argued otherwise  \cite{Achkar2014}.  Although quenched disorder can pin an otherwise slowly fluctuating order such that it is detectable by static probes, the stripe dynamics is expected to become glassy \cite{MVojta2009}.  In that case, the apparently static short-range CO configurations, or domain structures, correspond to long-lived metastable states.  However, metastable states may emerge even in the absence of disorder, as a result of frustration \cite{Schmalian2000,Sami2015}.  Therefore, the key questions are the role of disorder and the nature of dynamics.  A time-domain approach, sensitive to charge, is thus needed to answer these questions.

We report a novel study of domain dynamics in cuprates, based on time-dependent, \textit{nonequilibrium} charge transport.  We focus on \lnsco\, in the regime \textit{across} the CO (and structural) transition \cite{Tranquada1995,Tranquada1996,Zimmermann1998,Ichikawa2000}, where fluctuations are expected to be most pronounced.  In general, fluctuating order is characterized by correlations on short enough time and length scales, where the system is critical  \cite{MVojta2009}.  Using two different nonequilibrium protocols, we find evidence for metastable states and criticality.  The collective dynamics in the critical regime, i.e. of the fluctuating order, indicates strong pinning by disorder.  Surprisingly, these nonequilibrium effects are observed only when the critical region is approached from the low-temperature charge-ordered phase, strongly suggesting that the dynamics of the observed fluctuating order reflects that of the CO.

\textit{Material}.--- La$_{2-x-y}$RE$_y$Sr$_x$CuO$_4$ (RE = Nd or Eu) and La$_{2-x}$Ba$_x$CuO$_4$ undergo a structural transition at $T=T_{\textrm{LTT}}$ from the low-temperature orthorhombic (LTO) to a low-temperature tetragonal (LTT) phase, with the transition consisting of a $\phi=45^{\circ}$ rotation of the tilting axis of the oxygen octahedra surrounding the Cu atoms \cite{Hucker2012}.  In the LTT phase ($T<T_{\textrm{LTT}}$), the CuO$_6$ octahedra tilt about axes parallel to the Cu-O bonds, leading to two inequivalent Cu-O-Cu bonds or  anisotropy within the CuO$_2$ planes that stabilizes stripes.  The tilt axis and thus stripes are rotated by 90$^{\circ}$ from one CuO$_2$ layer to next \cite{Tranquada1995}.  In the basic picture, when both static spin and charge stripes are well developed at low enough $T<T_{\textrm{CO}}, T_{\textrm{SO}}$ ($T_{\textrm{CO}}$ and $T_{\textrm{SO}}$ are the onset of static charge and spin orders, respectively), doped holes populate antiphase domain walls, which run along the Cu-O bond direction and separate antiferromagnetic (AF) domains of spins on Cu sites (spin stripes).  However, slowing down of the charge dynamics and short-range CO have been observed already at $T>T_{\textrm{CO}}$ \cite{Hunt2001}.  
The LTO-LTT transition is manifested as a jump in $R(T)$, more pronounced in the out-of-plane than in the in-plane transport, and apparently enhanced when  $T_{\textrm{CO}}\sim T_{\textrm{LTT}}$ \cite{Nakamura1992,Tranquada1996,Takeshita2004}.  Thus we study the $c$-axis resistance $R_c$ 
%
\begin{figure*}
\includegraphics[width=17.8cm]{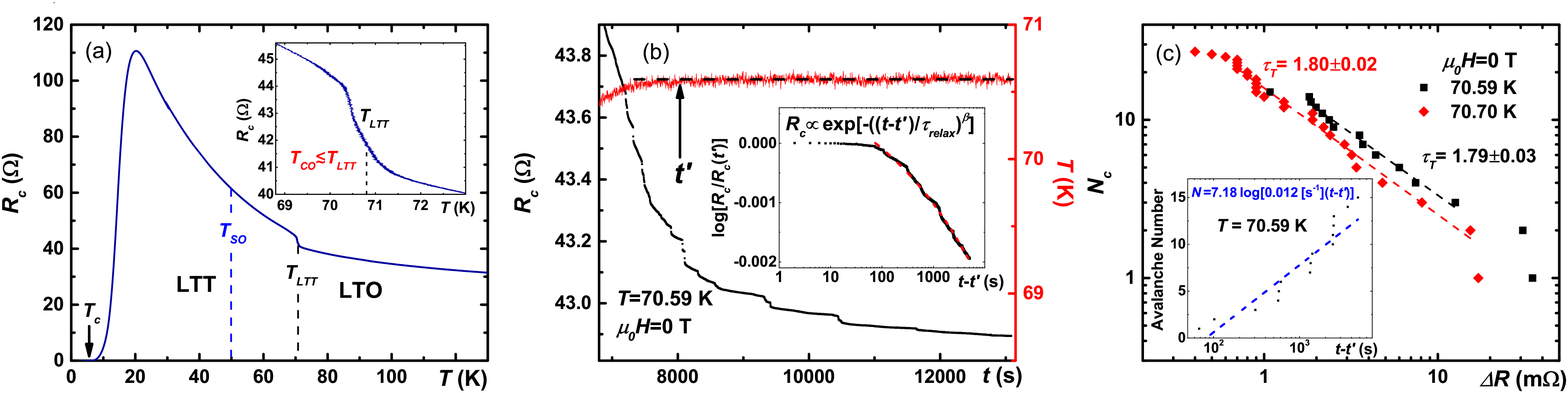}
\caption{(a) $R_c$ vs. $T$.  $T_c$ and $T_{\textrm{LTT}}$ were determined as shown; superconducting fluctuations vanish at $T\sim 30$~K \cite{Adachi2005,Xiang2008,Xie2011}.  $T_{\textrm{CO}}$ and $T_{\textrm{SO}}$ are from Refs.~\cite{Tranquada1996,Zimmermann1998,Ichikawa2000}.  Inset: Transition region.  (b) Relaxation and avalanches in $R_c$ after warming from $\sim 40$~K to $T=70.59$~K.  $T$ becomes stable at $t=t'$, but $R_c$ relaxes for hours after $t'$.  Inset: The relaxation, with avalanches removed, fitted to a stretched exponential function (dashed line), with $\beta=0.2$ and $\tau_{relax}=32$~s.  (c) The cumulative distribution $N_c$ of avalanche sizes $\Delta R$ for the data in (b) ($T=70.59$~K) and at $T=70.7$~K, obtained from ten separate measurements \cite{SM}.  Dashed lines are fits $N_c\sim (\Delta R)^{-(\tau_{T}-1)}$ with $\tau_{T}\approx 1.8$, as shown.  Inset: The occurrence of avalanches in (b) after $t'$; the dashed line is a logarithmic fit.  
\label{TempDependence}}
\end{figure*}
%
in \lnsco~(LNSCO), in which the apparent static charge ordering occurs at $T=T_{\textrm{CO}}\lesssim T_{\textrm{LTT}}\simeq 70$~K; $T_{\textrm{SO}}\simeq 50$~K \cite{Tranquada1996,Zimmermann1998,Ichikawa2000}.  For this doping, the LTO-LTT transition region is characterized by the presence of an intermediate, low-temperature less-orthorhombic (LTLO) phase, in which the rotation of the octahedral tilt axis is not complete, i.e. $0^{\circ}<\phi<45^{\circ}$ \cite{Crawford1991,Sakita1999}.  

\textit{Methods}.--- The LNSCO single crystal was grown by the traveling-solvent floating-zone technique. Detailed $R_c$ measurements were performed on a bar-shaped sample with dimensions $1.46\times 0.41\times 0.24$~mm$^3$ using a dc reversal technique \cite{keithley2008} (excitation current $I_{\textrm{exc}}=100~\mu$A in the Ohmic regime) \cite{SM}.  Magnetoresistance (MR) was measured in magnetic fields $H\parallel c$ up to 12~T, with sweep rates (0.05-0.5~T/min) that had no effect on our results \cite{SM}.  The sample becomes superconducting ($R_c=0$) at $T_c=(3.51\pm0.06)$~K, and  $T_{\textrm{LTT}}=(70.8\pm0.5)$~K is taken at the center of the jump in $R_c(T)$ [Fig.~\ref{TempDependence}(a)].  The width of the jump reflects the presence of the LTLO region \cite{Sakita1999}, which is quite narrow ($\sim 1$~K) in our crystal (cf. $\sim10$~K in Ref.~\cite{Sakita1999}), suggesting that the amount of excess oxygen is very small \cite{Hucker2004}.  A jump in $R_c$ is known to be accompanied by a small thermal hysteresis, which is attributed to the first-order nature of the structural transition \cite{Nakamura1992}. 
 
The  stability of the CO in cuprates implies that, if any metastable states are present, thermal fluctuations are too small to overcome the energy barriers that separate them.  Thus we study the response of the system to external perturbations \cite{SM}: (i) a change in $T$, and (ii) applied $H$.  All measurements were done with a controlled history.  Nonequilibrium effects were observed within the transition region [Fig.~\ref{TempDependence}(a) inset], as described below.

\textit{Response to a temperature change}.--- In the ``cooling'' procedure, the sample was cooled (0.08--0.1~K/min) from a high $T\gg T_{\textrm{LTT}} \simeq T_{\textrm{CO}}$, typically $\sim 90$~K, to the measurement temperature.  No intrinsic relaxations of $R_c$ with time $t$ were observed after cooling.  In the ``warming'' procedure, the sample was cooled from $\sim 90$~K down to $T\ll T_{\textrm{LTT}}\simeq T_{\textrm{CO}}$, typically $\sim 40$~K, but the results did not depend on the precise value of this $T$ or on the subsequent heating rate (0.07--1.4~K/min) to the measurement temperature \cite{SM}.  In that case, a striking difference was found compared to cooling: $R_c$ continued to relax for hours after the measurement temperature became stable [Fig.~\ref{TempDependence}(b)].  In addition, the relaxations were accompanied by avalanchelike jumps [Fig.~\ref{TempDependence}(b)] that occurred roughly with a logarithmic $t$ dependence (Fig.~\ref{TempDependence}(c) inset).  If avalanches are ``removed'' by shifting parts of the $R_{c}(t)$ curve by the size of the jumps, $\Delta R$, then the overall relaxation is described best with a stretched exponential, $R_{c}\propto\exp\{-[(t-t')/\tau_{relax}]^{\beta}\}$, with $\beta=0.2$ and $\tau_{relax}=32$~s (Fig.~\ref{TempDependence}(b) inset).  At short times, the relaxation is slower, but the range of the data is not sufficiently large to determine its form precisely.  Importantly, nonexponential relaxations are typical signatures of systems with many metastable states, as they reflect the existence of a broad distribution of relaxation times \cite{SM}.

%
\begin{figure*}
\includegraphics[width=17.8cm,clip=]{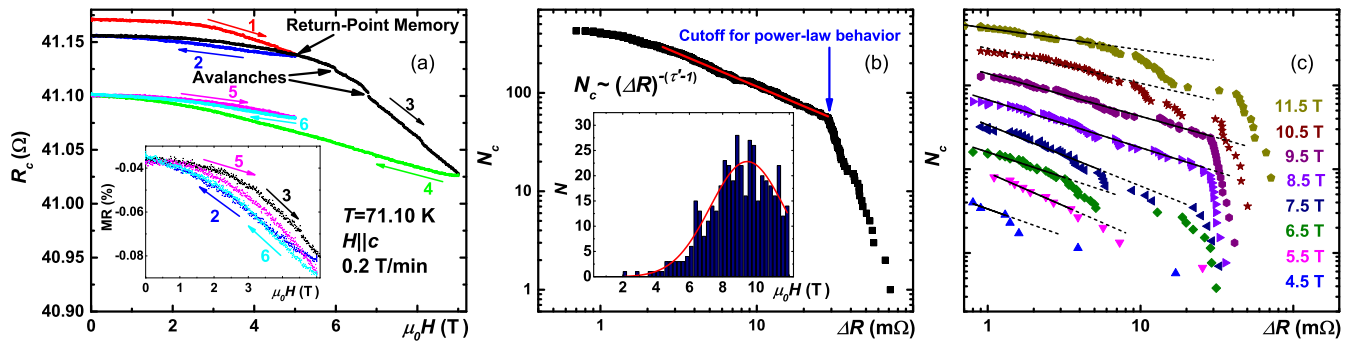}
\caption{(a) $R_c$ vs. $H$ after warming to $T=71.10$~K.  The arrows and numbers show the direction and the order of field sweeps.  The MR exhibits return-point memory, e.g. $R_c$ at 5~T has the same value after the first subloop $(2-3)$ has closed as in the initial sweep.  Inset: Subloops shifted vertically for comparison.  (b) $N_{c}(\Delta R)$ at $T=70.7$~K \cite{SM}.  The solid line is a fit $N_c\sim (\Delta R)^{-(\tau'-1)}$ with $\tau'=1.666\pm0.002$.  Inset: The number of avalanches ($N$) vs. $H$ (centers of 0.25~T-wide bins).  The solid line is a Gaussian fit, centered at $H_{c}=(9.5\pm 0.1)$~T with a 2.2~T half width.  (c) $N_{c}(\Delta R)$ for $T=70.7$~K and fixed $H$ (1~T bins), as shown.  Traces are offset vertically for clarity.  Solid lines are power-law fits; dashed lines are extrapolations of those fits.  At $H=H_c$, $N_c\sim (\Delta R)^{-(\tau-1)}$ with $\tau=1.54\pm 0.04$.
\label{Stats}}
\end{figure*}
%

Avalanches are collective rearrangements that occur as the system shifts from one metastable state to another \cite{Crackling-review}.  The distribution of avalanche sizes was analyzed by constructing the cumulative distribution $N_c(\Delta R)$ [Fig.~\ref{TempDependence}(c)], which  describes the probability that an avalanche has a size greater than or equal to $\Delta R$ \cite{Newman2005}.  A criterion proportional to the background standard deviation was used as the metric for counting avalanches as spikes in the derivative \cite{Spasojevic1996}.  We find a power-law dependence $N_c\sim (\Delta R)^{-(\tau_{T}-1)}$ ($\tau_{T}\approx 1.8$), indicating a broad range of scales, as expected in the critical regime. 

Thermal relaxations were observed between $\sim69.4$~K and $\sim71.5$~K.  At long enough times, they could no longer be detected, suggesting that thermal fluctuations were too small to overcome the energy barriers between metastable states.  In such cases, however, the energy barriers can be overcome by applying an external field, which modifies the free-energy landscape, driving the system from one metastable state to another.  Thus the MR was measured after the relaxations were no longer visible.  

\textit{Response to applied $H$.}--- As $T\rightarrow T_{\textrm{LTT}}^{+}$, we find that a weak, positive MR is replaced by the onset of negative MR, which grows with decreasing $T$ deep into the CO state \cite{SM}.  The properties of the negative MR [Fig.~\ref{Stats}(a)] suggest that $H\parallel c$ drives the system towards the lower-resistance LTO phase.  Indeed, a hysteresis [Fig.~\ref{Stats}(a)], observed in the transition region [Fig.~\ref{TempDependence}(a) inset], 
is a general feature of driven, athermal first-order phase transitions in the presence of disorder \cite{Sethna1993,Vives1994,Vives2001,Crackling-review,LeDoussal2009}.  Avalanches are observed during $H$ sweeps [Fig.~\ref{Stats}(a)], but \textit{only} with sweeps to $H$ higher than those applied previously.  Moreover, no avalanches have been seen in the MR obtained after cooling, although the MR still exhibits a hysteresis \cite{SM}.  Thus the occurrence of avalanches is \textit{asymmetric}: in both $H=0$ [Fig.~\ref{TempDependence}(b)] and the MR [Fig.~\ref{Stats}(a)] they are seen only when the system evolves from the CO/LTT phase.  The asymmetric avalanche behavior is uncommon; except for a couple of examples of related behavior \cite{Lilly1993,Wu1995,Adams2014}, avalanche distribution in various systems is symmetric for both cooling and heating through the transition, including structural martensitic transitions \cite{Vives-martensitic}, and across both branches of the hysteresis loop \cite{Sethna1993,Vives1994,Vives2001,Crackling-review,LeDoussal2009}.  On the other hand, some type of asymmetric behavior was seen in the thermal response of 1$T$-TaS$_2$ \cite{TaS2}, a conventional 
charge-density-wave (CDW) system, and attributed to the presence of metastable states.  Thus the observed asymmetry suggests the possibility that the presence of CO domains in the LTT phase in LNSCO may produce an additional manifold of metastable states that are not caused by disorder \cite{Sami2015}.  
 
The MR demonstrates return-point memory [Fig.~\ref{Stats}(a)] and a slight incongruence of closed subloops obtained between the same $H$ end points but with a different history (Fig.~\ref{Stats}(a) inset), indicating weak interactions between domains \cite{Bertotti}.  Although the MR hysteresis was observed previously near the LTO-LTT transition in a similar material \cite{Xu2000}, La$_{1.4-x}$Nd$_{0.6}$Sr$_{x}$CuO$_{4}$ with $x=0.10$ and 0.15, the novel evidence for interactions (incongruent subloops and avalanches) and return-point memory impose strong constraints on theory.  For example, the $T=0$ random-field Ising model (RFIM) exhibits all of the observed properties \cite{Sethna1993}, albeit no asymmetry.

To gather sufficient statistics, MR was measured ten times at $T=70.7$~K \cite{SM}.  Figure~\ref{Stats}(b) inset shows that the avalanches are observed only above a threshold field of $\lesssim 2$~T, suggesting that this is the minimum depinning field for the domains.  Indeed, the MR is  reversible for sweeps up to $H<2$~T.  Furthermore, the avalanches have a Gaussian field distribution, centered at $H_c= (9.5\pm 0.1)$~T.  This Gaussian distribution is consistent with the standard theoretical assumptions (e.g. Refs.~\cite{Sethna1993,Crackling-review}) about the distribution of local critical fields, due to disorder, around $H_c$, the critical driving field of the clean system.  We find that $N_c\sim (\Delta R)^{-(\tau'-1)}$, up to an exponential cut-off characteristic of the presence of strong pinning or disorder (e.g. Ref.~\cite{LeDoussal2009_2}).  Indeed, the exponent $\tau'=1.666\pm0.002$ and the existence of slow relaxations [Fig.~\ref{TempDependence}(b)] are consistent with a model for a three-dimensional (3D) system with strong pinning of many small domains, as opposed to the motion and depinning of large domain walls in the case of weak pinning \cite{Tadic2002}.  However, estimating the spatial size of the domains from $\Delta R$ may not be straightforward \cite{Sharoni2008}.

While the cut-off in $N_c(\Delta R)$ is related to the maximum size of the domains due to strong pinning, the power-law behavior reveals the existence of criticality in the system \cite{Perkovic1999,Vives2001}.  Strictly speaking, criticality is expected to occur only at a certain field.  Indeed, $N_c$ determined for fixed $H$ [Fig.~\ref{Stats}(c)] demonstrates that the power law is obeyed best for $H_c=9.5$~T, where $N_c\sim (\Delta R)^{-(\tau-1)}$ with $\tau=1.54\pm 0.04$.  Away from $H_c$, the distributions increasingly deviate from the power law, as expected.  The cut-off, however, does not vanish at $H_c$, indicating that the strength of disorder $W$ exceeds the critical level of disorder $W_c$ above which an infinite avalanche never occurs \cite{Sethna1993,Crackling-review}.  Nevertheless, $H_c(W)$ is clearly close enough to the critical point $H_{c}(W_{c})$ to observe the power law.  Hence, our results are consistent with the existence of an $H$-driven first-order phase transition with disorder.  Even the result for $\tau$ is consistent with general predictions \cite{Vives2001}, including that of the 3D RFIM, but further studies beyond the scope of this work are needed to identify the precise universality class of this transition.  We note though that the asymmetric occurrence of avalanches in the MR is in contrast to standard predictions \cite{Sethna1993,Vives1994,Vives2001,Crackling-review,LeDoussal2009}.

The avalanche occurrence as a function of $T$ was studied by calculating the average number of avalanches per MR sweep in a given, 0.1~K-wide temperature bin.  The distribution 
%
\begin{figure}
\includegraphics[width=8.5cm]{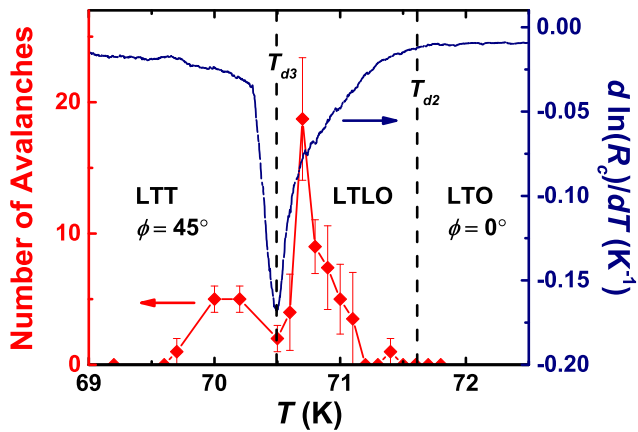}
\caption{The average number of avalanches per MR measurement (red diamonds), after warming, vs. $T$.  To increase the statistics, i.e. the number of data sets used, only avalanches for $H\leq 9$~T were included.  $d\ln R_c/dT$ (blue line) has a minimum at the suppression of avalanche occurrence.  The features of $d\ln R_c/dT$ marked by the vertical dashed lines correspond to $T_{d3}$ and $T_{d2}$, the temperatures of the LTT-LTLO and LTLO-LTO transitions, respectively \cite{Sakita1999}.  
\label{TempN}}
\end{figure}
%
has a large peak near the center of the $R_{c}(T)$ jump [$T_{LTT}$ in Fig.~\ref{TempDependence}(a)],  i.e. in the LTLO region, with a second, smaller peak in the LTT phase, with no avalanches observed in the LTO structure (Fig.~\ref{TempN}).  In general, fluctuations are expected to peak at the phase transition.  For example, in La$_{2-x}$Sr$_x$NiO$_4$, charge stripe fluctuations were found to peak just above or around $T_{\textrm{CO}}$, and to vanish gradually at higher and lower $T$ \cite{Anissimova2014}.  Thus we tentatively attribute the two peaks in Fig.~\ref{TempN} to the onset of CO and precursor nematic order \cite{Kivelson2003,Achkar2016,Pelc2016} in the LTT and LTLO regions, respectively.

\textit{Discussion}.--- Clearly, the resistance measurement is a direct, bulk probe of the charge degrees of freedom, capable of testing their dynamics on exceptionally long time scales.  Thus the evidence obtained from $R_c$ for metastable states, correlated behavior, and criticality, and only when the critical region is approached from the charge-ordered phase, strongly suggests that it reflects the dynamics of CO.  It is thus interesting to speculate about a possible relationship between the power-law distribution of avalanche sizes found here to that of the spatial distribution of charge ``stripes'' below $T_{\textrm{CO}}$ in HgBa$_2$CuO$_{4+y}$ at optimal doping \cite{Campi2015}, and about the similarities of our results in LNSCO to the dynamical behavior of conventional pinned CDW systems.  For example, the latter are well known to exhibit nonexponential relaxations and history-dependent effects as evidence for metastable states \cite{Monceau-review}.  Likewise, our evidence for the $H$-driven transition bears some similarities to the RFIM universality class of the nematic transition in the presence of (weak) disorder \cite{Nie2014}.  Random local critical fields in LNSCO are likely due to the tendency of Nd$^{3+}$ moments towards ferromagnetic ordering in $H\parallel c$ \cite{Xu2000}.

Since CO in LNSCO is intricately related to LTT distortions \cite{Tranquada1995,Tranquada1996,Zimmermann1998,Ichikawa2000}, we consider the possibility that the observed domain dynamics reflect changes in the crystal structure.  We first note that, since the asymmetric avalanche behavior is uncommon, it does not seem plausible that it would occur at the transition between very similar phases, such as LTT, LTLO, and LTO.  Additional, important insight can be gained from the $c$-axis MR measurements in $H\parallel c$ on a striped La$_{1.7}$Eu$_{0.2}$Sr$_{0.1}$CuO$_4$, in which the structural and CO transitions are well separated ($T_{\textrm{LTT}}\simeq 126$~K; $T_{\textrm{CO}}\sim 40$~K) \cite{Fink2011,Autore2014,Klauss2000}.  Here the MR remains weak and increasingly \textit{positive} as $T$ is reduced through $T_{\textrm{LTT}}$ \cite{SM}, indicating that $H$ cannot drive the structural transition.  In contrast, as $T$ is decreased further, a \textit{negative} MR sets in at $T\sim T_{\textrm{CO}}$, and continues to grow with decreasing $T$ deep into the CO state \cite{SM}. Hence, the evidence for the $H$-driven first-order phase transition and the associated nonequilibrium dynamics obtained from the negative MR in LNSCO seems related to the CO and not to the structural transition.  The onset of the negative MR is, in fact, revealed as a signature of the onset of CO, but the precise mechanism for the negative MR remains unclear.  In particular, there are no proposed spin-based mechanisms relevant to the regime studied, i.e. for $T>T_{\textrm{SO}}$, to the best of our knowledge.  On the other hand, the applied $H$ may affect CO by coupling to the orbital motion \cite{Monceau-review}.  Indeed, the magnetic length ($\sim 8$~nm) at $H\sim 10$~T is comparable to the stripe correlation length $\sim 11$~nm in LNSCO \cite{Wilkins2011}, but other orbital mechanisms might be also at play \cite{Baledent2010}.  Similar studies on a stripe-ordered La$_{2-x}$Ba$_x$CuO$_4$ might provide additional insight into this issue.

We have reported the detection and study of the fluctuating order across the CO transition in \lnsco.  A picture emerges of interacting domains that are trapped in long-lived metastable states and strongly pinned by disorder.  Avalanches, which occur in response to external perturbations, represent \textit{collective} rearrangements of the domains.  The surprising asymmetry of the observed nonequilibrium effects, however, suggests the presence of an additional manifold of metastable states in the CO phase that are not caused by disorder.  These results set important qualitative and quantitative constraints on theories of dynamic stripes, and also clarify the conditions necessary for the observation of dynamic, as opposed to static, domains: a sufficiently large external perturbation and  measurements with a well-controlled history, pointing a way to detecting fluctuating domains in the cuprates using also other experimental techniques.  Nonequilibrium protocols in charge transport can be extended to other correlated-electron systems, such as iron pnictides, to probe charge domain dynamics.

\begin{acknowledgments}
We thank P. Kuhns, A. Reyes, and Z. Shi for experimental assistance, B. Tadi\'c, V. Dobrosavljevi\'c, O. Vafek, A. Chubukov, and E. Manousakis for discussions, NSF Grants No. DMR-1307075 and No. DMR-1707785, and the NHMFL through the NSF Cooperative Agreement No. DMR-1157490 and the State of Florida for financial support.  This work  was performed in part at the Aspen Center for Physics, which is supported by NSF grant PHY-1066293.
\end{acknowledgments}

\clearpage

\setcounter{figure}{0}
\makeatletter
\makeatletter \renewcommand{\fnum@figure}{{\figurename~S\thefigure}}
\makeatother

\onecolumngrid{

\begin{center}
\large\textbf{Supplemental Material for: Collective Dynamics and Strong Pinning Near the Onset of Charge Order in \lnsco}
\vspace{12pt}

\normalsize

P.~G.~Baity,$^1$ T. Sasagawa,$^2$ and Dragana Popovi\'c$^1$
\vspace{6pt}

\small
$^1$\textit{National High Magnetic Field Laboratory and Department of Physics,\\ Florida State University, Tallahassee, Florida 32306, USA}\\
$^2$\textit{Materials and Structures Laboratory, Tokyo Institute of Technology, Kanagawa 226-8503, Japan}

\end{center}
}

\twocolumngrid

\setlength{\textfloatsep}{5pt plus 1.0pt minus 2.0pt}
\textit{Methods}.--- The electrical contacts on the sample were made by evaporating Au and annealing in air at 700$^{\circ}$C.  

The magnetic field sweep rates (0.05-0.5~T/min) were low enough to avoid heating of the samples from eddy currents.   In this range, the results did not depend on the sweep rate.

A common technique for probing the dynamics of CDWs in quasi-one-dimensional conductors, or conventional CDW systems, is the application of a very high electric field to study the nonlinear conductivity [S1].  However, similar attempts to detect collective stripe motion in cuprates [S2] and nickelates [S3, S4], which employed electric fields up to $10^{3}$~V/cm, found only nonlinear transport effects that could be attributed to the conventional Joule heating.  In contrast, our study focuses on the Ohmic transport, using a small electric field $\sim 0.06$~V/cm.  Studies of the Ohmic conductivity in conventional CDW systems have revealed evidence for metastable states, such as nonexponential relaxations [S1].  Therefore, we use a low electric field to measure the response of the Ohmic conductivity to two other external perturbations, in particular, to changes in temperature and magnetic field.

\textit{Response to a temperature change}.--- For technical reasons, the cooling rate was limited to 0.08--0.1~K/min, while warming rates were varied from 0.07~K/min to 1.4~K/min.  Within this range, the results did not depend on the heating rate.  Furthermore, it was not possible to use higher warming rates, because they led to the temperature overshoot while stabilizing the temperature, so that relaxations and avalanches were then not observed.

Since the relaxations observed after warming up to the transition region are nonexponential (see Fig. 1(b) inset of the main text), there is no single, characteristic lifetime associated with the dynamics in the charge-ordered phase.  Furthermore, we do not find any temperature dependence of the stretched exponential relaxation, i.e. of $\tau_{relax}$ or $\beta$, within our experimental accuracy.

\begin{figure}[h]
\includegraphics[width=8.0cm,clip=]{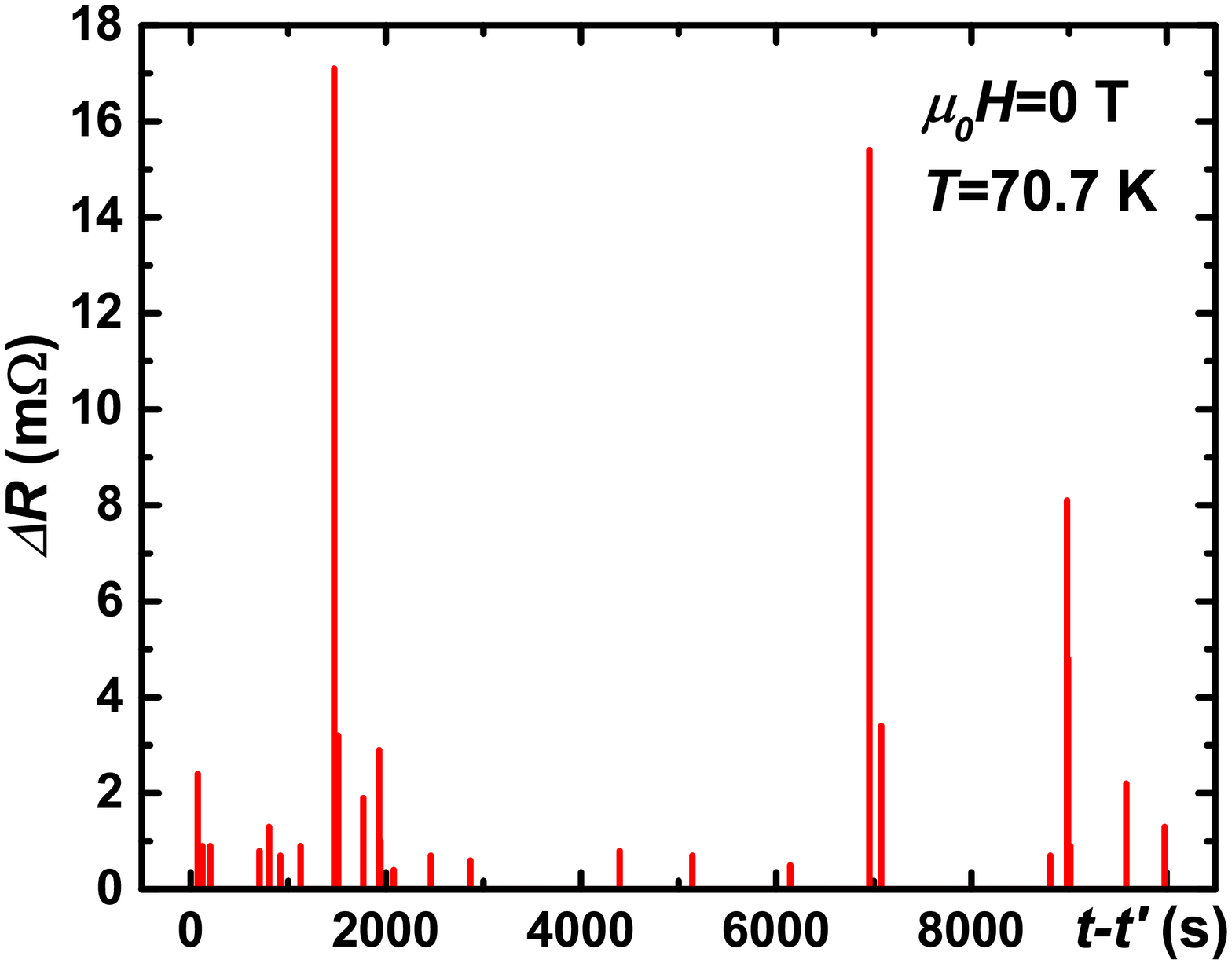}
\caption{Avalanche sizes $\Delta R$ vs. acquisition time ($t'$ is the time when the temperature becomes stable), obtained from ten separate measurements after warming to $T=70.7$~K.
\label{fig:sizes}}
\end{figure}

Figure~S\ref{fig:sizes} shows the avalanche sizes $\Delta R$ as a function of the 
%
%
%
\begin{figure*}
\includegraphics[width=17.0cm,clip=]{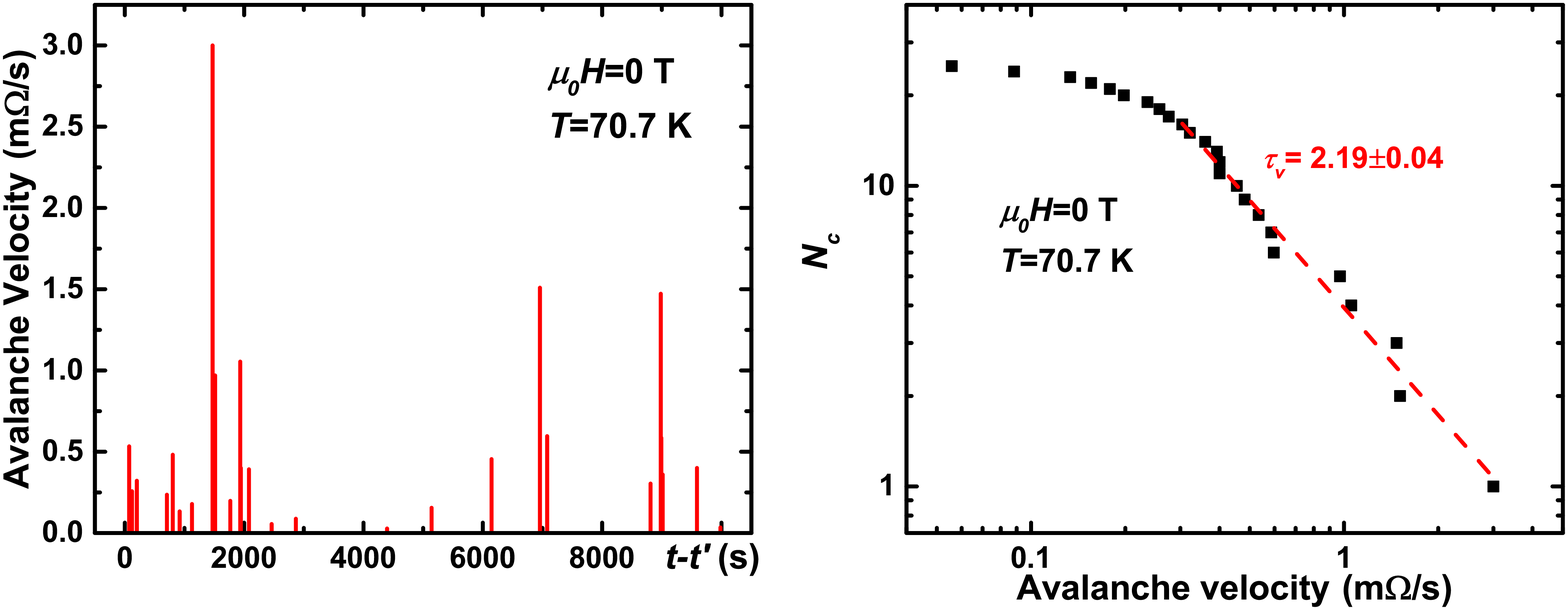}
\caption{Left: Avalanche velocity, $\Delta R/\Delta t$, vs. acquisition time ($\Delta t$ is the avalanche duration and $t'$ is the time when the temperature becomes stable), obtained from ten separate measurements after warming to $T=70.7$~K.  Right: The cumulative distribution of the data in the left panel.  The dashed line is a fit $N_c\sim (\Delta R/\Delta t)^{-(\tau_{v}-1)}$ with $\tau_{v}\approx 2.2$, as shown.
\label{fig:velocities}}
\end{figure*}
%
acquisition time, obtained after warming to $T=70.7$~K.  The time series looks like crackling noise [S5], i.e. it exhibits discrete events with different sizes.  Indeed, the cumulative distribution of avalanche sizes obeys a power law, $N_c\sim (\Delta R)^{-(\tau_{T}-1)}$ with $\tau_{T}\approx 1.8$ (red diamonds in Fig.~1(c) of the main text), confirming a broad, scale-free  distribution of sizes.  Similar behavior was observed in a charge-stripe state of a nickelate La$_2$NiO$_{4.14}$ using low electric fields [S4], albeit with $\tau_{T}\approx 1.3$ and after cooling.  Furthermore, no asymmetry in the occurrence of avalanches was reported.
\vspace{1pt}
\newline
\indent The ``velocity'' of each avalanche was calculated by dividing its size $\Delta R$ by its duration $\Delta t$.  The time series of $\Delta R/\Delta t$ vs. acquisition time (Fig.~S\ref{fig:velocities}, left) also resembles crackling noise, and the corresponding cumulative distribution obeys a power law $N_c\sim (\Delta R/\Delta t)^{-(\tau_{v}-1)}$ with $\tau_{v}\approx 2.2$ (Fig.~S\ref{fig:velocities}, right).
\vspace{1pt}
\newline
\indent \textit{Response to applied $H$.}--- Figure~S\ref{fig:LNSCO-MR} reveals, for $H\parallel c$, a weak positive magnetoresistance (MR) at $T>T_{\textrm{LTT}}$ and the onset of the negative MR as $T_{\textrm{CO}}\lesssim T_{\textrm{LTT}}\simeq 71$~K is approached.  The negative MR grows with decreasing $T$, deep into the charge-ordered phase.  The positive MR found at $T>T_{\textrm{LTT}}$ does not exhibit any hysteresis.  The rest of the discussion is focused on the negative MR observed in the transition region [e.g.~Fig.~1(a) inset], where nonequilibrium phenomena are observed.
\vspace{1pt}
\newline
\indent Figure~S\ref{CoolingMR} demonstrates that the negative MR curves obtained after cooling do not show any avalanches even though the MR still exhibits a hysteresis.
\vspace{1pt}
\newline
\indent Figure~S\ref{TenMR} shows ten MR measurements obtained after warming to the same $T$ to gather sufficient statistics on avalanches.  The power-law distribution of avalanche sizes (Fig. 2(b) of the main text) implies also the existence of many small avalanches that cannot be resolved within the noise of our experiment.  We find that the avalanches that are observable within our experimental resolution contribute about 40\% of the MR measured at 12~T (Fig.~S\ref{TenMR}).
\vspace{1pt}
\newline
\indent The cumulative distribution of the velocities of avalanches shown in Fig.~S\ref{TenMR} also obeys a power law, $N_c\sim (\Delta R/\Delta t)^{-(\tau_{v}'-1)}$ with $\tau_{v}'\approx 2.0$ (Fig.~S\ref{fig:velocity-H}).  
\vspace{1pt}
\newline
\indent For completeness, we note that, for $H\parallel ab$, the MR in the transition region is positive and hysteretic [S6], but with no avalanches (not shown).  $H\parallel ab$ thus appears to favor the CO/LTT phase.
\vspace{1pt}
\newline
\indent \textit{Discussion}--- For comparison, some out-of-plane MR measurements were performed on a La$_{1.7}$Eu$_{0.2}$Sr$_{0.1}$CuO$_4$ single crystal, also grown by the traveling-solvent floating-zone technique [S7].  The samples were bar-shaped with dimensions $0.41\times 0.34\times 1.67$~mm$^3$ and $1.20\times 0.52\times 1.61$~mm$^3$.  For this material, $T_{\textrm{LTT}}\simeq 126$~K, $T_{\textrm{CO}}\sim 40$~K, and $T_{\textrm{SO}}\sim 15$~K [S8-S10]; evidence for the electronic nematicity has not been reported.  Figure~S\ref{fig:LESCO-MR}(a) shows that a weak, positive MR is observed as $T$ is reduced through the structural transition.  The positive MR grows with decreasing $T$, deeper into the LTT phase.  However, this behavior is reversed as $T_{\textrm{CO}}$ is approached [Fig.~S\ref{fig:LESCO-MR}(b)]: at $T\sim T_{\textrm{CO}}$, we observe the onset of the negative MR, which grows with decreasing $T$, deeper into the charge-ordered phase.

\vspace{6pt}
\noindent\textbf{{References}}
\newline
\noindent [S1] For a review, see P.~Monceau, Advances in Physics \textbf{61}, 325 (2012).
\vspace{1pt}
\newline
\noindent [S2] A.~N.~Lavrov, I.~Tsukada, and Y.~Ando, Phys.~Rev.~B \textbf{68}, 094506 (2003).
\vspace{1pt}
\newline
\noindent [S3] M.~H\"ucker, M.~v.~Zimmermann, and G.~D.~Gu, Phys.~Rev.~B \textbf{75}, 041103(R) (2007).
\vspace{1pt}
\newline
\noindent [S4] A.~Pautrat, F.~Giovannelli, and N.~Poirot, Phys.~Rev.~B \textbf{75}, 125106 (2007).
\vspace{1pt}
\newline
\noindent [S5] J.~P.~Sethna, K.~A.~Dahmen, and C.~R.~Myers, Nature \textbf{410}, 242 (2001).
\vspace{1pt}
\newline
\noindent [S6] Z. A. Xu, N. P. Ong, T. Noda, H. Eisaki, and S. Uchida, Europhys. Lett. {\bf 50}, 796 (2000).
\vspace{1pt}
\newline
\noindent [S7] N.~Takeshita, T.~Sasagawa, T.~Sugioka, Y.~Tokura, and H.~Takagi, J. Phys. Soc. Jpn. \textbf{73}, 1123 (2004).
\vspace{1pt}
\newline
\noindent [S8] J. Fink, V. Soltwisch, J. Geck, E. Schierle, E. Weschke, and B. B\"uchner, Phys. Rev. B {\bf 83}, 092503 (2011).
\vspace{1pt}
\newline
\noindent [S9] M. Autore, P. Di Pietro, P. Calvani, U. Schade, S. Pyon, T. Takayama, H. Takagi, and S. Lupi, Phys. Rev. B {\bf 90}, 035102 (2014).
\vspace{1pt}
\newline
\noindent [S10] H.-H. Klauss, W. Wagener, M. Hillberg, W. Kopmann, H. Walf, F. J. Litterst, M. H\"ucker, and B. B\"uchner, Phys. Rev. Lett. {\bf 85}, 4590 (2000).
%
\begin{figure}[h]
\includegraphics[width=7.8cm,clip=]{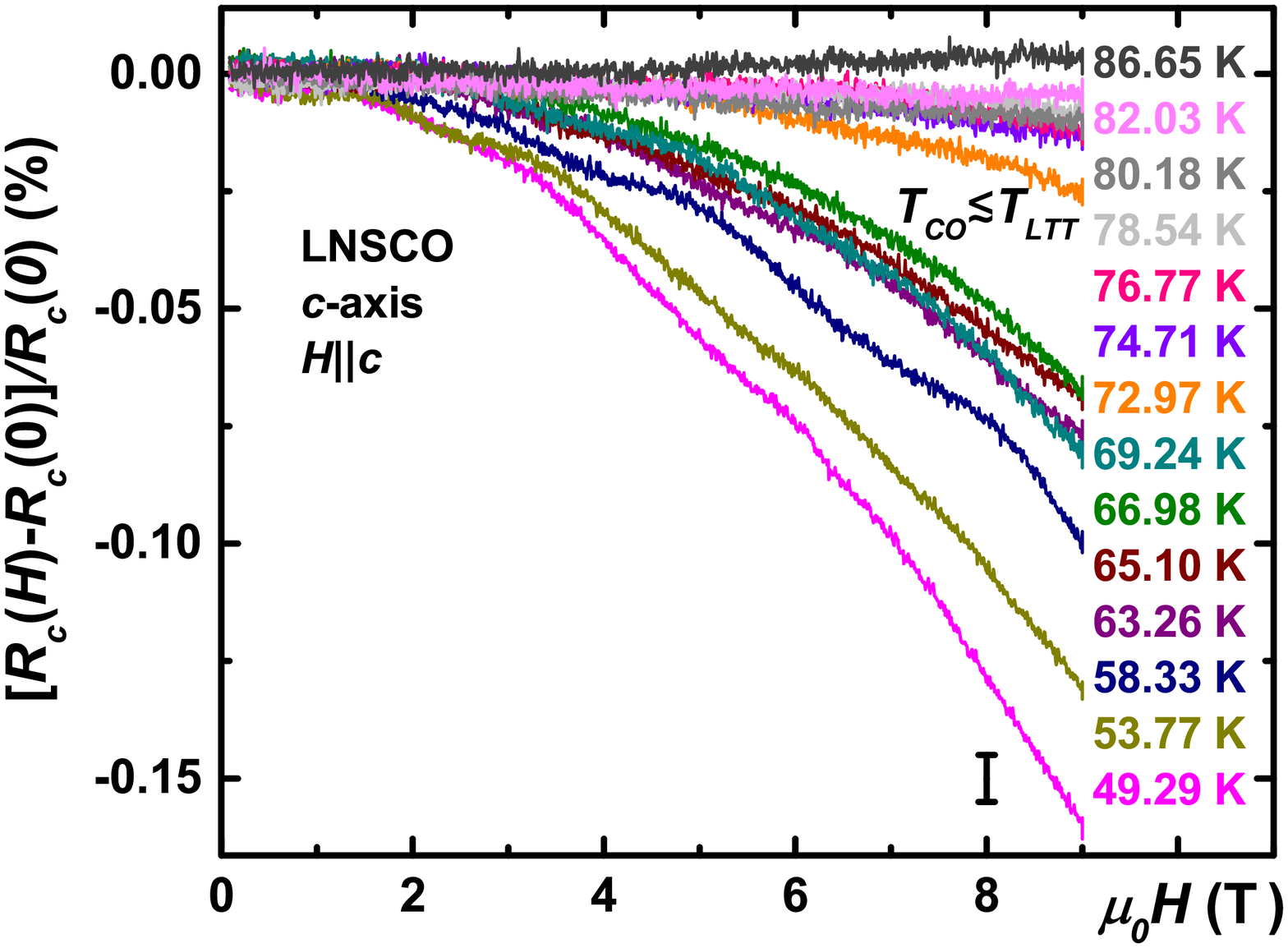}
\caption{MR for several $T$ above and well below $T_{\textrm{CO}}\lesssim T_{\textrm{LTT}}\simeq 71$~K.  The error bar results from the uncertainty due to temperature fluctuations.
\label{fig:LNSCO-MR}}
\end{figure}
%
%
\begin{figure}
\includegraphics[width=8.0cm,clip=]{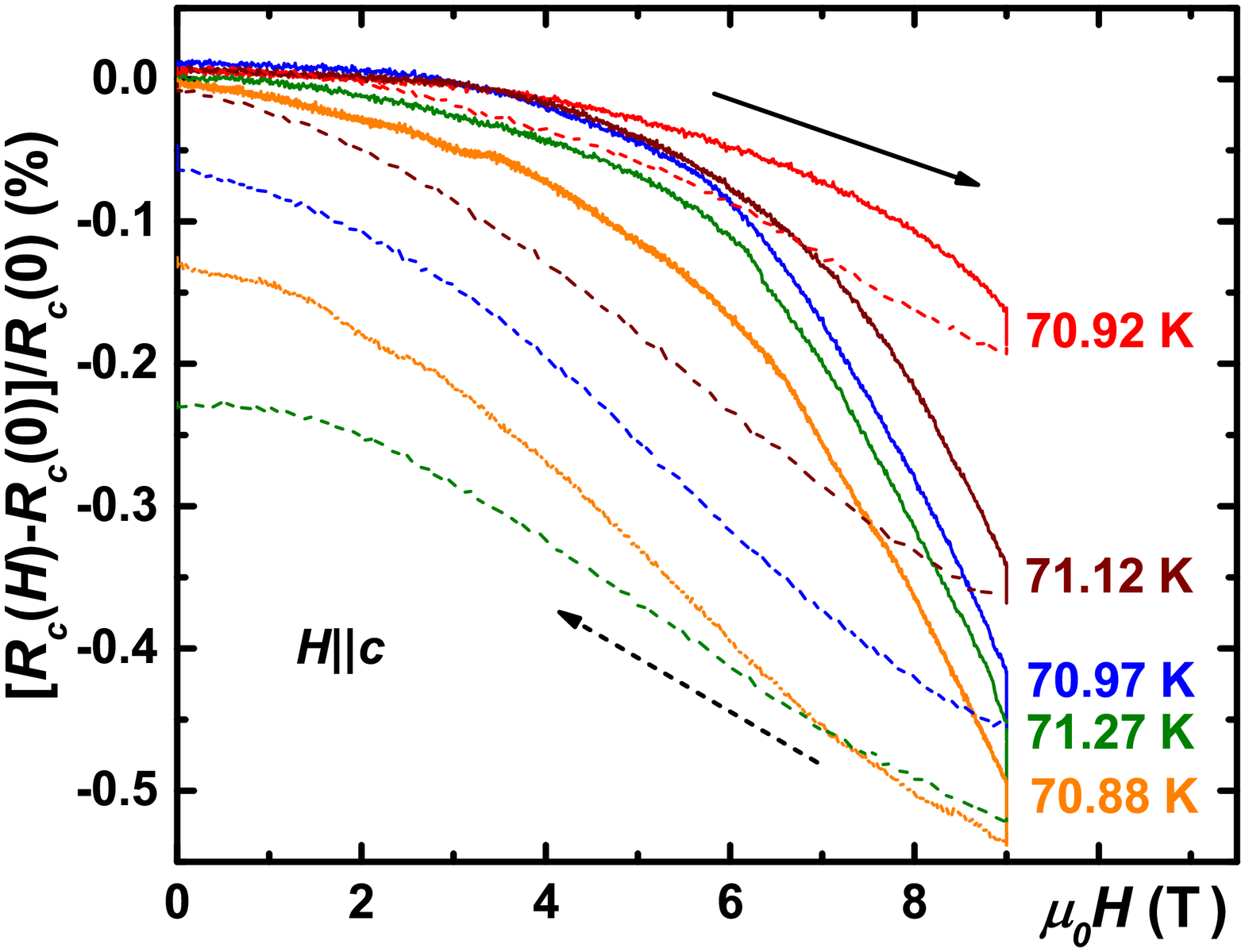}
\caption{MR for several $T$ within the charge-order and structural transition regime, as shown.  Each MR was measured after cooling the sample from 90~K, i.e. as the transition was approached from above or the LTO phase.  The MR curves exhibit hysteresis, but no avalanches.  The arrows denote field sweep direction.  The field sweep rates were 0.1~T/min at $T=70.88$~K, 0.25~T/min at 71.12~K, 0.3~T/min at 70.97~K, and 0.4~T/min at 70.92~K and 71.27~K.
\label{CoolingMR}}
%
\vspace{30pt}
%
\includegraphics[width=8.0cm,clip=]{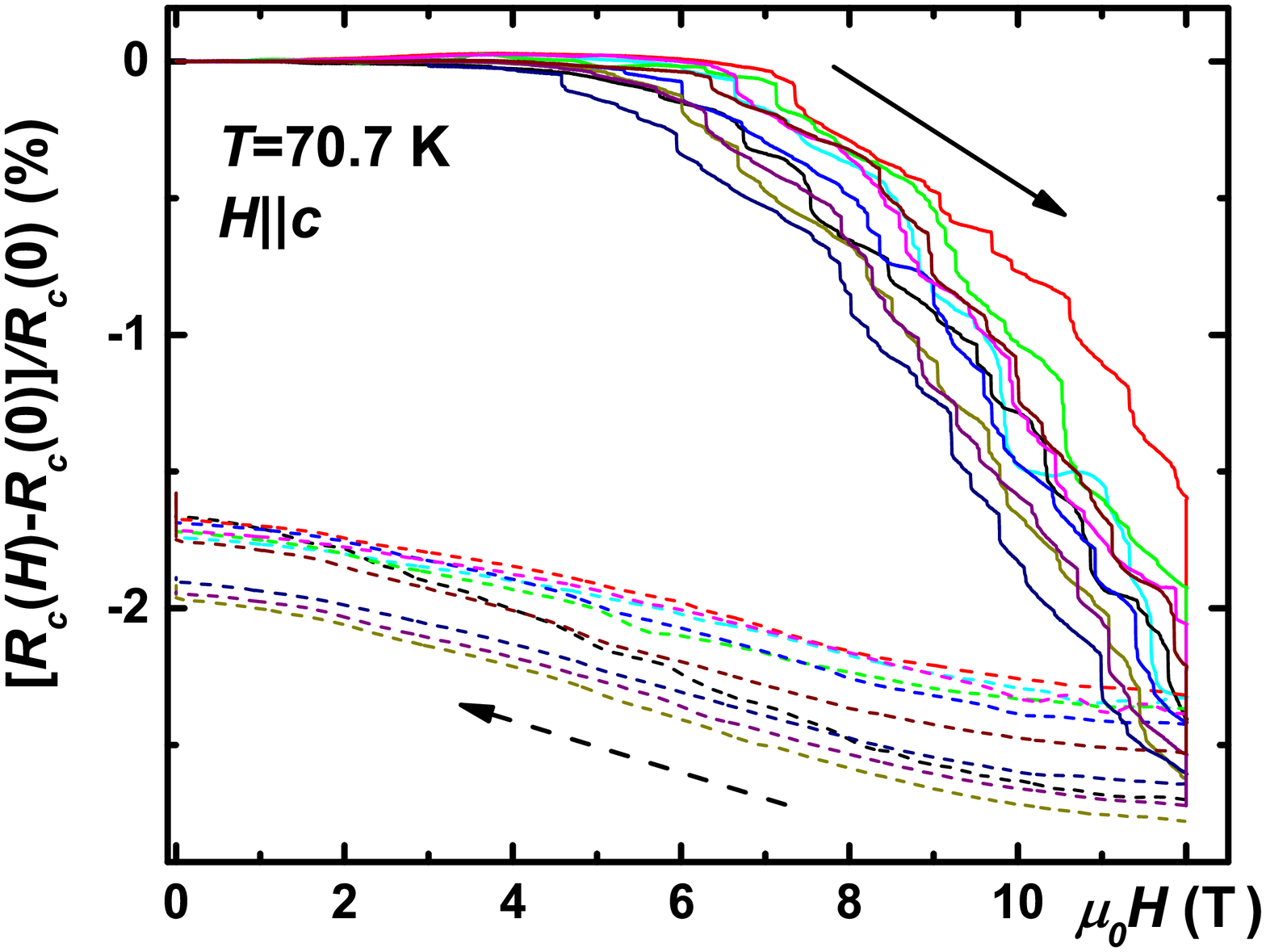}
\caption{Ten MR curves, each obtained after warming from $\sim 40$~K to $T=70.7$~K.  The field sweep rates were 0.1~T/min.  The arrows show the direction of field sweeps.  Avalanches are only observed in the initial sweep.  
\label{TenMR}}
\end{figure}
%
%
\begin{figure}[H]
\includegraphics[width=8.0cm,clip=]{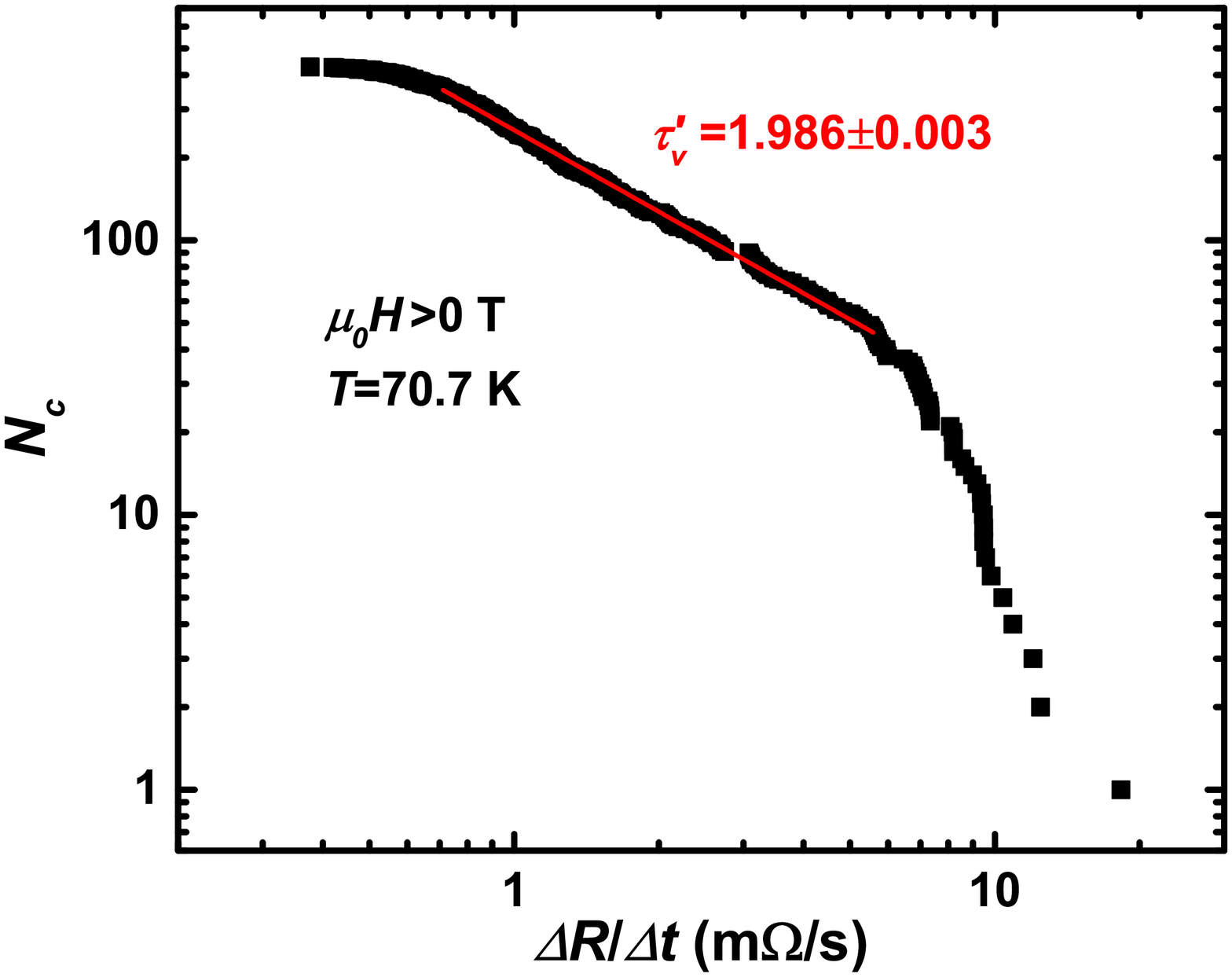}
\caption{The cumulative distribution of the velocities of avalanches shown in Fig.~S\ref{TenMR}.  The dashed line is a fit $N_c\sim (\Delta R/\Delta t)^{-(\tau_{v}'-1)}$ with $\tau_{v}'\approx 2.0$, as shown. 
\label{fig:velocity-H}}
%
\vspace{20pt}
%
\includegraphics[width=8.0cm,clip=]{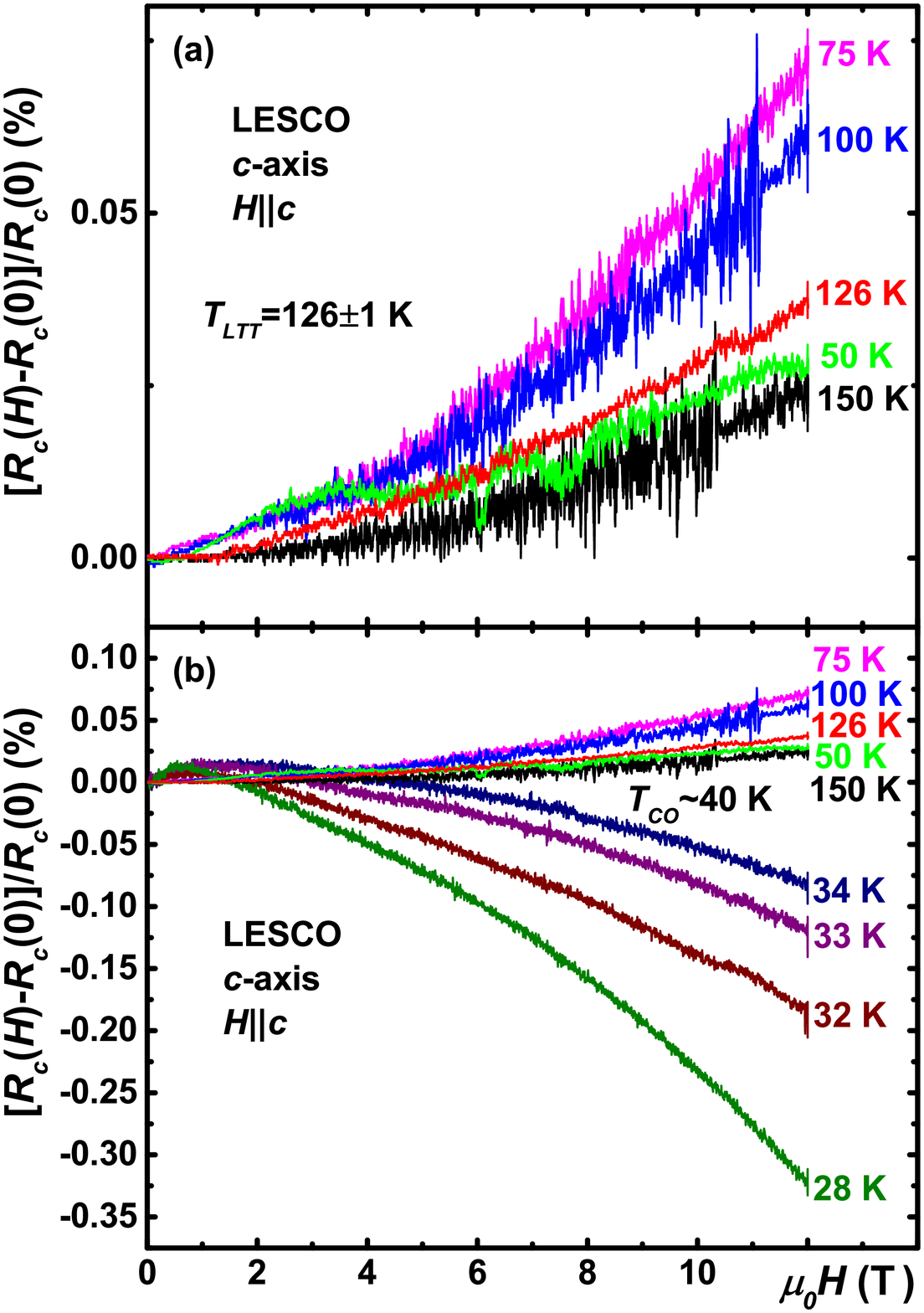}
\caption{MR of La$_{1.7}$Eu$_{0.2}$Sr$_{0.1}$CuO$_4$ for several $T$ (a) near the structural transition at $T_{\textrm{LTT}}\simeq 126$~K, and (b) also through the charge-order transition at $T_{\textrm{CO}}\sim 40$~K.
\label{fig:LESCO-MR}}
\end{figure}
%

\end{document}